\documentclass[letterpaper]{article} 
\usepackage{aaai25}  
\usepackage{graphicx} 
\usepackage{natbib}  

\usepackage{mathtools}  
\usepackage{amssymb} 
\usepackage{caption} 
\usepackage{tikz}
\usepackage{numprint}
\usetikzlibrary{matrix,backgrounds}
\usetikzlibrary{shapes.geometric}
\usetikzlibrary{calc,shadows}
\usepackage{caption}

\usepackage{graphicx} 

\usepackage{newfloat}

\usepackage{listings}
\DeclareCaptionStyle{ruled}{labelfont=normalfont,labelsep=colon,strut=off} 
\usepackage[linesnumbered,vlined,algoruled]{algorithm2e}
\usepackage{graphicx}
\usepackage{textcomp}
\usepackage{xcolor}
\def\BibTeX{{\rm B\kern-.05em{\sc i\kern-.025em b}\kern-.08em
T\kern-.1667em\lower.7ex\hbox{E}\kern-.125emX}}

\usepackage{cleveref}

\newtheorem{definition}{Definition}
\newtheorem{theorem}{Theorem}
\begin{document}

    \title{Solving Multi-Agent Multi-Goal Path Finding Problems in Polynomial Time}

\author{Stefan Edelkamp \\ 
Charles University \\
Prague, Czech Republic \\
\texttt{stefan.edelkamp@mff.ktiml.cuni.cz}
}
    \maketitle

   \begin{abstract}
In this paper, we plan missions for a fleet of agents in undirected graphs, such as grids, with multiple goals. In contrast to regular multi-agent path-finding, the solver finds and updates the assignment of goals to the agents on its own. In the continuous case for a point agent with motions in the Euclidean plane, the problem can be solved arbitrarily close to optimal. For discrete variants that incur node and edge conflicts, we show that it can be solved in polynomial time, which is unexpected, since traditional vehicle routing on general graphs is NP-hard. We implement a corresponding planner that finds conflict-free optimized routes for the agents. Global assignment strategies greatly reduce the number of conflicts, with the remaining ones resolved by elaborating on the concept of ants-on-the-stick, by solving local assignment problems, by interleaving agent paths, and by kicking agents that have already arrived out of their destinations. 
        
        
    \end{abstract}

\section{Introduction}

\begin{figure}[h!]
    \centering
\begin{small}
\begin{minipage}{2.5cm} 
\begin{tikzpicture}[scale=0.76]
    \draw[line width=0.04cm] (1,0) grid (4,5);

    \node[draw,circle,fill=black!90,inner sep=3pt] at (1,5) {}; 
    \node[draw,circle,fill=black!90,inner sep=3pt] at (1,2) {}; 
    \node[draw,circle,fill=black!90,inner sep=3pt] at (3,4) {}; 
    \node[draw,circle,fill=black!90,inner sep=3pt] at (2,2) {}; 
    \node[draw,circle,fill=black!90,inner sep=3pt] at (4,3) {}; 

    \node[draw,circle,fill=yellow!90,inner sep=2pt] at (1,1) {}; 
    \node[draw,circle,fill=purple!90,inner sep=2pt] at (3,3) {}; 
    \node[draw,circle,fill=green!90,inner sep=2pt] at (1,3) {}; 
    
    \node[draw,star,fill=red!90,inner sep=3pt] at (4,1) {}; 
    \node[draw,star,fill=red!90,inner sep=3pt] at (1,0) {}; 
    \node[draw,star,fill=red!90,inner sep=3pt] at (3,5) {}; 

    \node[above right] at (1,1) {Agent 1};
    \node[above] at (3,3) {Agent 2};
    \node[below right] at (1,3) {Agent 3};
    \node[below left] at (4,1) {Goal 1};
    \node[below right] at (1,0) {Goal 2};
    \node[below right] at (3,5) {Goal 3};
    
    \draw[line width=0.09cm,->,yellow] (1,1) -- (2,1) -- (3,1) -- (4,1) ; 
    \draw[line width=0.09cm,->,green] (1,3) -- (2,4) -- (3,5); 
    \draw[line width=0.09cm,->,purple] (3,3) -- (3,2) -- (2,1) -- (1,0); 
\end{tikzpicture}  
\end{minipage}
\begin{minipage}{2.5cm} 
\begin{tikzpicture}[scale=0.76]
    \draw[line width=0.04cm] (1,0) grid (4,5);

    \node[draw,circle,fill=black!90,inner sep=3pt] at (1,5) {}; 
    \node[draw,circle,fill=black!90,inner sep=3pt] at (1,2) {}; 
    \node[draw,circle,fill=black!90,inner sep=3pt] at (3,4) {}; 
    \node[draw,circle,fill=black!90,inner sep=3pt] at (2,2) {}; 
    \node[draw,circle,fill=black!90,inner sep=3pt] at (4,3) {}; 
    
    \node[draw,circle,fill=yellow!90,inner sep=2pt] at (1,1) {}; 
    \node[draw,circle,fill=purple!90,inner sep=2pt] at (3,3) {}; 
    \node[draw,circle,fill=green!90,inner sep=2pt] at (1,3) {}; 
    
    \node[draw,star,fill=red!90,inner sep=3pt] at (4,1) {}; 
    \node[draw,star,fill=red!90,inner sep=3pt] at (1,0) {}; 
    \node[draw,star,fill=red!90,inner sep=3pt] at (3,5) {}; 

    \node[above right] at (1,1) {Agent 1};
    \node[above] at (3,3) {Agent 2};
    \node[below right] at (1,3) {Agent 3};
    \node[below left] at (4,1) {Goal};
    \node[below right] at (1,0) {Goal};
    \node[below right] at (3,5) {Goal};
    
    \draw[line width=0.09cm,->,yellow] (1,1) -- (1,0) ; 
    \draw[line width=0.09cm,->,green] (1,3) -- (2,4) -- (3,5); 
    \draw[line width=0.09cm,->,purple] (3,3) -- (4,2) -- (4,1); 
\end{tikzpicture}  
\end{minipage}
\begin{minipage}{2.5cm} 
\begin{tikzpicture}[scale=0.76]
    \draw[line width=0.04cm] (1,0) grid (4,5);

    \node[draw,circle,fill=black!90,inner sep=3pt] at (1,5) {}; 
    \node[draw,circle,fill=black!90,inner sep=3pt] at (1,2) {}; 
    \node[draw,circle,fill=black!90,inner sep=3pt] at (3,4) {}; 
    \node[draw,circle,fill=black!90,inner sep=3pt] at (2,2) {}; 
    \node[draw,circle,fill=black!90,inner sep=3pt] at (4,3) {}; 
    
    \node[draw,circle,fill=yellow!90,inner sep=2pt] at (1,1) {}; 
    \node[draw,circle,fill=green!90,inner sep=2pt] at (1,3) {}; 
    
    \node[draw,star,fill=red!90,inner sep=3pt] at (4,1) {}; 
    \node[draw,star,fill=red!90,inner sep=3pt] at (1,0) {}; 
    \node[draw,star,fill=red!90,inner sep=3pt] at (3,5) {}; 

    \node[below right] at (1,1) {Agent 1};
    \node[below right] at (1,3) {Agent 2};
    \node[below left] at (4,1) {Goal};
    \node[below right] at (1,0) {Goal};
    \node[below right] at (3,5) {Goal};
    
    \draw[line width=0.09cm,->,yellow] (1,1) -- (1,0) -- (2,0) -- (3,0) -- (4,1); 
    \draw[line width=0.09cm,->,green] (1,3) -- (2,4) -- (3,5); 
\end{tikzpicture}  
\end{minipage}
\end{small}    
   \caption{Regular, Anonymous, and Combinatorial MAPF (left to right). where the search graph nodes are line  intersections. red stars represent target nodes, obstacles are shown as gray circles, agents are colored dots and their travel is indicated by a chain of colored edges of the agent's color.
   For MAPF a fixed assignment from goals to agents is enforced, in anonymous MAPF there are as many goals as agents, while for CMAPF there might be more goals than agents. }   
    \label{fig:hungarian}
\end{figure}
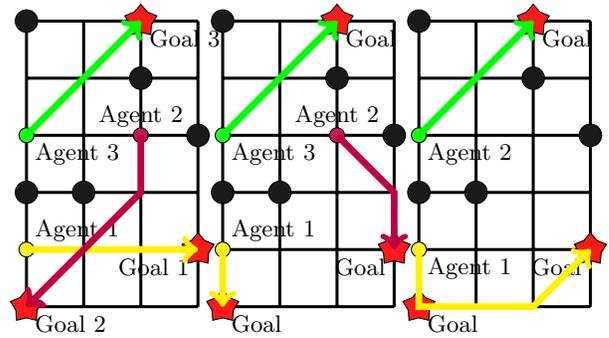

Multi-agent path finding (MAPF) is the problem of computing collision-free paths for a set of agents from their current locations to given destinations. Application examples include automated warehouse systems, office agents, and non-player moves in video games. 
Regular MAPF~\cite{DBLP:conf/socs/SternSFK0WLA0KB19} 
with a fixed assignment of $k$ agent locations to $k$ goals usually executed in an undirected and uniformly weighted graph such as a grid has impacted considerable research progress. 
Anonymous multi-agent path finding (AMAPF), also referred in some works as Unlabeled MAPF~\cite{Zain}, is MAPF where the goal assignment is left to the solver (see Figure~\ref{fig:hungarian}). For multiple goals to be assigned to each agent, this is called combinatorial multi-agent path finding (CMAPF).

Multiple goals are a significant extension for current MAPF solvers. For example, \cite{CBSS} study an extension of the conflict-based search (CBS) of MAPF to CMAPF, which was later shown to be non-optimal~\cite{DBLP:conf/socs/MouratidisNK24}. The core issue was the greedy manner of the chained A* algorithm, clearing the hash table of nodes at different stages of the search. Mixing individual A* search trees into one retains optimality, but such a composite search is less efficient.

Although polynomial-time algorithms even for makespan-optimal AMAPF are known~\cite{Zain,YuValle}, the resulting computational complexities are high, as algorithms usually act on the time-expanded graph, so that the input size of the network is already large, which together with the superlinear time complexity for network flow leads to considerable running times in practice~\cite{Zain,MaAAMAS16}.
In contrast to this line of work, in our approach we mainly work on the shortest-path reduced graph. We derive efficient implementations for pre-computing single-source (all-targets) shortest paths.

For AMAPF, we compute an optimal solution to the assignment problem. Historically, the Hungarian algorithm~\cite{5Hungarian} to solve the assignment problem initially took time $O(k^4)$, with $k$ being the number of agents on the graph. It was refined to $O(k^3)$, and among these cubic-time algorithms, we selected the method of shortest augmentation paths~\cite{jonker}. The original problem is to optimize the cumulative cost of the assignment, which corresponds to the sum of cost, while the variant that minimizes the maximum cost, called the bottleneck assignment problem, can also be solved in cubic time~\cite{burkard2012assignment}. 
When no conflicts between agents on the shortest paths occur, the solutions obtained from solving the assignment problem are already optimal for AMAPF. 

To address the tour-assignment problem for CMAPF, we employ a bounded number of iterations of the NRPA algorithm~\cite{Rosin} where every rollout yields a valid assignment (anytime behavior). Recall that NRPA is among the state-of-the-art solutions to vehicle routing problems~\cite{DBLP:journals/aicom/CazenaveLTK21}.
Given any set of agent tours, our proposed simulation-based conflict resolution runs in polynomial time and yields a collision-free schedule. The generated solution will be optimized and optimal in many cases. In other words, solving does not impose optimality but requires 
polynomial-time algorithms.

The paper is structured as follows. We briefly review related work on the problem. Next, we introduce the general solution principle that we apply. We consider the computational limits and possibilities for anonmymous and combinatorial MAPFs, and introduce polynomial-time solutions to these problems. Experimentally, we study the reduction obtained in the sum of costs and makespan, as well as the reduction and elimination of conflicts in the set of 500 well-known MAPF benchmark instances from the Moving AI repository \cite{stern2019mapf}. Finally, we reflect on the results obtained and indicate future research avenues.

\section{Related Work}

Various research has tried to solve travel salesman problems efficiently~\cite{tspsolve,LiangVRPMP, LKTSP}. 
Moreover, there is a huge research body on the efficient solution of MAPF variants, e.g. \cite{Expectmakespanopimal,  MAPFLN1,  MAPFLN2,PushSwap, 
PushRotate,  
decouple, 
multigoal, 
Guidance,  
Silver}.

Regular MAPF on undirected graphs has been solved~\cite{RogerH12}, referring to the tractable algorithms of
Wilson~\cite{wilson} for one blank position on biconnected graphs, and of Kornhauser, Miller and Spirakis~\cite{KornhauserMS84} for the general case. They did not attempt to keep the number of plan steps low, but 
required polynomial-time algorithms.

More recently, Regular MAPF on directed graphs was shown to be NP-complete~\cite{Nebel23}. Given that the problem was known to be NP-hard, Nebel proved that the problem is also contained in NP.
Another interesting MAPF variant is the option to
disconnect trailers or containers from vehicle agents~\cite{Bachor}.


From a logistics perspective, with CMAPF we consider extensions of the vehicle routing problem (VRP)~\cite{vrp} to indoor navigation, where the objective is to find optimized tours for a fleet of vehicles to service a given set of customer requests while avoiding collisions. Usually in logistics, the assigning of the goals to the vehicles has to be found by the solver. 
Dynamic VRPs were studied by~\cite{Bullo}, and vehicle routing for agents with temporal constraints by \cite{VRPMP}, using a high-level task planner to statically assign waypoints to vehicles and a low-level motion planner that generates a feasible path that respects the vehicle motion model.
%
An approach to MAPF with precedence-constrained goal sequences was studied by~\cite{precedenceconstraints}, while 
\cite{DBLP:journals/ai/AndreychukYSAS22} highlighted the applicability of multiagent pathfinding algorithms in real-world applications; and raised questions of how to discretize time effectively, proposing algorithms for finding optimal solutions that do not rely on any time discretization.

Although we have not found reduction-based optimal AMAPF or CMAPF solvers for sum-of-cost optimization (without agents disappearing at the goals), suboptimal AMAPF algorithms were proposed and empirically shown to provide promising solutions~\cite{Okumura}. A decentralized AMAPF was studied in~\cite{decentralized}. An AMAPF variant 
involving the coloring of teams was studied in~\cite{DBLP:conf/flairs/BartakIS21}. 
There is precurser work on target assignments for AMAPF~\cite{MaAAMAS16},
Its algorithmic approach is 
network flow and conflict-based search.

\section{Solution Principle}

\begin{figure}[t]
\centering
\begin{tikzpicture}[scale=0.7]
    \draw[thick, brown, line width=0.06cm] (1,0) -- (8,0);
    
    \node[below] at (2,-0.1) {$a_1$};
    \node[below] at (4,-0.1) {$a_2$};
    \node[below] at (7,-0.1) {$a_3$};
    \node[above] at (8.5,-0.1) {$a_4$};
    
    \draw[->, very thick] (2, 0) -- (2.5, 0); 
    \draw[->, very thick] (4, 0) -- (4.5, 0); 
    \draw[->, very thick] (7, 0) -- (6.5, 0); 
    \draw[->, very thick] (8.5, 0) -- (8.5, -0.5); 
    \node[draw, circle, fill=blue, inner sep=2pt] at (2, 0) {}; 
    \node[draw, circle, fill=red, inner sep=2pt] at (4, 0) {}; 
    \node[draw, circle, fill=green, inner sep=2pt] at (7, 0) {}; 
    \node[draw, circle, fill=purple, inner sep=2pt] at (8.5, 0) {}; 
\end{tikzpicture}

    \centering
    \caption{Ants-on-a-stick problem, exchanging direction on collision with one ant falling off at and of stick.}
   
    \label{fig:antproblem}
\end{figure}
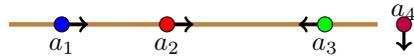

\begin{figure}[t]
    \centering
\begin{tikzpicture}[scale=1.2,
    node distance=1cm,
    ant node/.style={circle, draw, fill=red!90, inner sep=1pt, minimum size=1mm}, 
    box node/.style={draw, fill=brown!10, inner sep=1pt, minimum size=1mm}, 
    ant2 node/.style={circle, draw, fill=blue!90, inner sep=1pt, minimum size=1mm}, 
    ant3 node/.style={circle, draw, fill=green!90, inner sep=1pt, minimum size=1mm}, 
    path node/.style={circle, draw, fill=blue!20, inner sep=1pt, minimum size=1.2mm}, 
    edge style/.style={draw, thick, brown,opacity=1}, 
    path1/.style={red!100!, very thick}, 
    path2/.style={blue!100!, very thick}, 
    path3/.style={green!100!, very thick}, 
    ant/.style={
        decoration={markings, mark=at position #1 with {\arrow[scale=1.0]{Stealth}}},
        postaction={decorate}
    }
]

\node[path node] (A) at (0,0) {\tiny A};
\node[path node] (B) at (3,1) {\tiny B};
\node[path node] (C) at (2,-2) {\tiny C};
\node[path node] (D) at (5,0) {\tiny D};
\node[path node] (E) at (4,-3) {\tiny E};
\node[path node] (F) at (7,-1) {\tiny F};



\draw[edge style] (A) -- (B) node[midway, above] {$a_1$};
\draw[edge style] (A) -- (C) node[midway, left] {$a_2$};
\draw[edge style] (B) -- (D) node[midway, above right] {$a_3$};
\draw[edge style] (B) -- (C) node[midway, left] {};
\draw[edge style] (C) -- (D) node[midway, right] {$a_6$};
\draw[edge style] (C) -- (E) node[midway, below] {$a_4$};
\draw[edge style] (D) -- (F) node[midway, above] {};
\draw[edge style] (E) -- (F) node[midway, above] {$a_5$};
\draw[edge style] (D) -- (E) node[midway, right] {};

\draw[path1,opacity=0.8,->] (F) -- (E);
\draw[path1,opacity=0.8,->] (E) -- (C);
\draw[path1,opacity=0.8,->] (C) -- (A);

\draw[path2,opacity=0.8,->] (A) -- (B);
\draw[path2,opacity=0.8,->] (B) -- (D);
\draw[path2,opacity=0.8,->>] (D) -- (F);

\draw[path3,opacity=0.8,->] (A) -- (C);
\draw[path3,opacity=0.8,->] (C) -- (D);
\draw[path3,opacity=0.8,->] (D) -- (F);

\node[box node] (anta1) at ($(C)!0.75!(A)$) {\tiny C};
\node[ant3 node] (ant1) at ($(C)!0.7!(A)$) {};
\node[box node] (anta2) at ($(E)!0.23!(C)$) {\tiny C,A};
\node[ant node] (ant2) at ($(E)!0.33!(C)$) {};
\node[box node] (anta3) at ($(F)!0.23!(E)$) {\tiny E,C,A};
\node[ant node] (ant3) at ($(F)!0.3!(E)$) {};

\node[box node] (anta4) at ($(A)!0.306!(B)$) {\tiny B,D,F};
\node[ant2 node] (ant4) at ($(A)!0.4!(B)$) {};
\node[box node] (anta5) at ($(B)!0.50!(D)$) {\tiny D,F};
\node[ant2 node] (ant5) at ($(B)!0.6!(D)$) {};

\node[box node] (anta6) at ($(C)!0.63!(D)$) {\tiny D,F};
\node[ant3 node] (ant6) at ($(C)!0.7!(D)$) {};
\node[box node] (anta7) at ($(C)!0.25!(A)$) {\tiny A};
\node[ant node] (ant7) at ($(C)!0.3!(A)$) {};

    \draw[shorten >=0.7cm,shorten <=0cm, ->, very thick] (ant6) -- (D); 
    \draw[shorten >=0.5cm,shorten <=0cm, ->, very thick] (ant5) -- (D); 
    \draw[shorten >=1.6cm,shorten <=0cm, ->, very thick] (ant4) -- (B); 
    \draw[shorten >=2.3cm,shorten <=0cm, ->, very thick] (ant3) -- (E); 
    \draw[shorten >=1.15cm,shorten <=0cm, ->, very thick] (ant2) -- (C); 
    \draw[shorten >=1.65cm,shorten <=0cm, ->, very thick] (ant1) -- (C); 
    \draw[shorten >=1.65cm,shorten <=0cm, ->, thick] (ant7) -- (A); 

\end{tikzpicture}
\caption{Agents-on-a-graph problem with agents following their individual shortest-paths, while exchanging goal agendas (that are attached to the agents) on collision. 
}
 
\label{fig:antsongraph}
\end{figure}
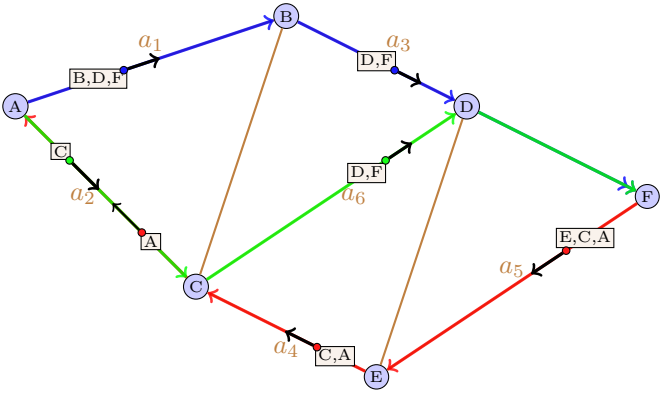

Our approach to solving the problem is based on the following intriguing mathematical puzzle (see Figure~\ref{fig:antproblem}
):
Several a(ge)nts are dropped on a $1m$ stick. Each a(ge)nt travels to the left or right with a constant speed $1m$ / minute. When two a(ge)nts meet, they bounce off each other and reverse direction. When an a(ge)nt reaches an end of the stick, it falls. At some point all the a(ge)nts will have fallen. The time at which this occurs will depend on the initial configuration of the a(ge)nts. We observe that two a(ge)nts bouncing off each other is equivalent to two a(ge)nts that pass through each other, instead of turning and exchanging their intended direction. In this way, all the a(ge)nts fall after traversing the length of the stick and it will be empty after 1 minute.

 We exploit the solution principle with agents moving on individual shortest path towards their targets, while exchanging goals agendas on every
collision (see Figure~\ref{fig:antsongraph}).
Assuming point agents, we observe the following result.

\begin{theorem}[Polytime Constant-Approximation]
  For any given tour assignment of $k$ infinitely small agents traveling from their starting locations to a set of goals in linear motion along the shortest-path edges in an undirected graph embedded in the Euclidean plane, let the total (maximal) travel time $L$ correspond to the sum (max) of the travel time over the edges accumulated for all agents. For any fixed value of $\epsilon >0$, the agents never use more than time $L +\epsilon$, while the collision resolution runs in polynomial time.
\end{theorem}

{\bf Proof.} Meeting on graph edges, agents simply swap their directions and their goal agendas, as illustrated in the ants-on-a-stick and agents-on-a-graph problems. If one agent has terminated at its destination and is on the way of some other agent, it is kicked out of it for the incoming agent to take its place (a process that we call cuckoo'ing). Assume now that $l>1$ agents meet at the same time at a junction $j$. When they are infinitesimally close to $j$ we reassign their goal agendas, with a resolution found by calling the cubic-time assignment problem solver for these agents on their respective next goal. In this way, all agents remain on the set of precomputed shortest paths to the goals. If agents aim to travel in the same direction along the same edge, we interleave them (indicated as zipping in Figure~\ref{fig:zipping}). By construction of the continuous problem for infinitesimal small agents, there is always space between two agents for a third one, so that we let one agent wait for a short time at the cost of at most $\epsilon/(kn)$, where $k$ is the number of agents and $n$ is the number of graph nodes. Even if all $k$ agents would meet at every junction, the accumulated delay does not exceed the time $\epsilon$. Hence, the obtained solution will be at most of length $L + \epsilon$. With at most a polynomial number of collisions, the total time complexity is polynomial.  \hfill $\Box$ \\

Let \emph{sum-of-cost} denote the sum of all travel and \emph{makespan} the maximum of travel cost over the set of agents. By applying polynomial time optimal algorithms for the initial assignment problems, we have derived a polynomial-time algorithm that yield solutions arbitrarily close to optimal for the continuous version of AMAPF. However, as solving the vehicle routing problem
is NP-hard, we cannot warrant
polynomial time for optimal solving a  continuous version CMAPF on general undirected graphs.


\section{Polynomial-Time Solutions}

In an undirected graph with $n$ nodes and a set of $k$ moving agents, we consider a multi-goal multi-agent path-finding problem. For a fixed assignment from $k$ agents to $k$ goals, we have the (nonoptimal) Regular MAPF problem that has been solved in polynomial time for an undirected graph~\cite{RogerH12,KornhauserMS84}. The work goes back to \cite{wilson}, who proved this for $k=n-1$.


\subsection{Anonymous MAPF}

For the Anonymous MAPF problem with $k$ agents and $k$ goals and makespan optimization, there are optimal polynomial algorithms~\cite{flow,Zain}.
The main idea is to replicate the underlying problem graph along the time line (adding a new layer for each set of waiting action) and looking for an optimal flow of the network. 
The network-flow approach does not easily carry over to multiple goals --- our main focus. However, for the sake of readability, we start with AMAPF.
 
\begin{definition}[Discrete Anonymous MAPF] 
Let $G=(V,E,w)$ be a weighted graph with $w: E \rightarrow \mathbb{N}$. 
We assume that the graph is undirected, so that for all $(u,v) \in E$ 
we have $(v,u) \in E$ and $w(u,v) =w(v,u)$. Edge weights must be polynomially bounded and we include self-loop edges for waiting at each node with cost 1. In the 
\emph{Asynchronous MAPF problem}, \emph{AMAPF} for short, for a set of agents $R$ with $k=|R|$ 
and pairs start and goal locations 
collision-free routes $\pi_i$, 
$i \in \{1,\ldots,k\}$, have to be found 
that for each agent starts in $s_i$, ends in some goal location and that together
minimize accumulated travel weight 
$\sum_{i=0}^k \sum_{(u,v) \in \pi_{i}} w(u,v)$ 
(sum-of-cost) or
maximum travel time 
$\max_{i=0}^k \sum_{(u,v) \in \pi_{i}} w(u,v)$
(makespan).
\end{definition}

AMAPF can be efficiently solved non-optimally by finding an assignment with the Hungarian algorithm and then applying the pebble motion solver of~\cite{RogerH12,KornhauserMS84} to the resulting MAPF. In contrast, our approach is more dynamic, as we allow reassignments of goals during the solution process. A collision occurs during the simulation of the routes, if two agents are located
at a node at the same time (node conflict) or on traversing an edge at the same time (edge conflict), so that, in addition to agent moving actions, wait actions might
be needed. Any grid can be easily compiled into an unweighted graph, with each cell representing a node. 

To decide which actions of individual agent need to be completed in exact
order and when agents can go at their own pace, we maintain an
agent dependency graph, a directed tree graph,
where each node represents a move
that should be issued by the agents. The edges between them represent dependencies between the actions. When in the graph one agent $a_i$ depends on another agent $a_j$, it means that agent $a_i$ needs to wait if $a_j$ is forced to wait. This graph is updated in each simulation step. Usually, edges are graph edges pointing to occupied adjacent nodes, but the relation can be extended through goal areas by following shortest paths. 

For undirected graphs $G = (V,E,w)$ with $n = |V|$ and $e = |E|$ and integer weights, Thorup \cite{DBLP:conf/focs/Thorup97} proposed an involved shortest-path algorithm that runs in linear time $O(n+e)$.
For our purpose, applying Dijkstra's algorithm is fast enough.  
For any weighted graph finding the shortest-paths from a single node to all other nodes takes time $O(e+n \log n)$ (with Fibonacci heaps), s.t.\ for $k$ agents, computing (and storing) all shortest paths runs in at most $O(k(e+n \log n))$ time. The time reduces to $O(k(n \log n))$ for planar graphs as $e = O(n)$ by Euler's formula, whereas for 4-connected grid graphs, for each agent a linear-time breadth-first search suffices.

\begin{theorem} [Polytime AMAPF]
For $k$ agents in an undirected weighted graph $G=(V,E,w)$ AMAPF is solvable in polynomial time wrt.\ $n=|V|$ and $k$.
\end{theorem}

{\bf Proof.} The proof is constructive by proposing an algorithm. While solving the assignment problem runs on the compressed graph and is polynomial in $k$, the simulation of the solution and conflict resolution runs on the uncompressed graph and is polynomial in time $n$ and $k$.

As a first step, we compute the shortest-path reduction of the graph. Applying Dijkstra's algorithm for each agent by computing the shortest paths backwards from every goal yields polynomial time (for path tracking, we keep all the shortest paths in memory). 



As a second step, we compute the matrix of pairwise distances between all start to goal nodes, which ---by chaining back associated shortest-path pointers--- is available in polynomial time using the precomputed information on the shortest paths of the first step. 

As a third step, we compute an assignment of the agents' starting locations to their respective goals, weighted with shortest path distances, in cubic time, asking for a mapping where the total (or maximum) travel time is minimized.  

In the fourth step, we simulate the solution in the graph to handle collision conflicts. We look at the precomputed shortest-path distance table to find the next location to go to for each agent. For each agent, we denote its position and orientation (that is, the direction of the next node to visit) and maintain the agent-dependency graph of links from one agent to the next. 

We are using the \emph{agents-on-a-graph} metaphor to exchange the goal assignment (and subsequent shortest-path tables) among the agents on a potential conflict. 
If at least one agent moves at a time along any shortest path edge towards a goal, the total number of iterations of the simulation is bounded by the total path length of all shortest paths, which is polynomial.
Usually, the number of iterations will be bounded by the length of the longest of the shortest paths. 
Subtleties arise, as agents must not collide so that they may be forced to wait. We will show that collisions can be resolved in polynomial time. 

\begin{figure}[t]
    \centering
\begin{minipage}{2.5cm} 
\begin{tikzpicture}[scale=1]
    \draw[line width=0.04cm] (1,0) grid (3,4);

    \node[draw,circle,fill=red!90,inner sep=2pt] at (2,1) {}; 
    \node[draw,circle,fill=green!90,inner sep=2pt] at (2,3) {}; 

    
     \draw[line width=0.09cm,->,red] (2.1,1) -- (2.1,3) -- (3,4) ; 
    \draw[line width=0.09cm,->,green] (1.9,3) -- (1.9,0); 
\end{tikzpicture}  
\end{minipage}
\begin{minipage}{2.5cm} 
\begin{tikzpicture}[scale=1]
    \draw[line width=0.04cm] (1,0) grid (3,4);

    \node[draw,circle,fill=red!90,inner sep=2pt] at (2,1) {}; 
    \node[draw,circle,fill=green!90,inner sep=2pt] at (2,3) {}; 

    
     \draw[line width=0.09cm,->,red] (2,3) -- (3,4); 
    \draw[line width=0.09cm,->,green] (2,1) -- (2,0); 
\end{tikzpicture}  
\end{minipage}
   \caption{Agenda swap. For node conflicts of agents pointing on the same node but in the opposite direction, the agents exchange their shortest path tables and goal agendas. This option can be extended to resolving via a local assignment problems in fixed area around agents. }
   
    \label{fig:agenda}
\end{figure}
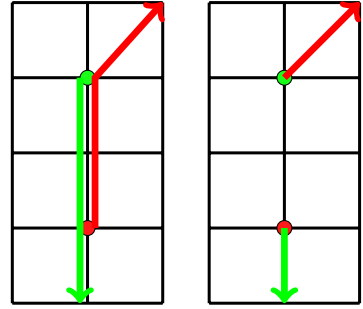

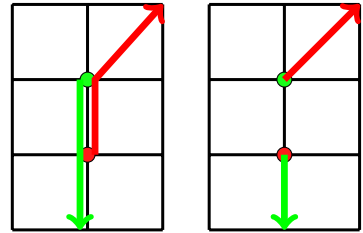
\begin{figure}[t!]
    \centering
\begin{minipage}{2.5cm} 
\begin{tikzpicture}[scale=1]
    \draw[line width=0.04cm] (1,0) grid (3,3);

    \node[draw,circle,fill=red!90,inner sep=2pt] at (2,1) {}; 
    \node[draw,circle,fill=green!90,inner sep=2pt] at (2,2) {}; 

    
     \draw[line width=0.09cm,->,red] (2.1,1) -- (2.1,2) -- (3,3); 
    \draw[line width=0.09cm,->,green] (1.9,2) -- (1.9,0); 
\end{tikzpicture}  
\end{minipage}
\begin{minipage}{2.5cm} 
\begin{tikzpicture}[scale=1]
    \draw[line width=0.04cm] (1,0) grid (3,3);

    \node[draw,circle,fill=red!90,inner sep=2pt] at (2,1) {}; 
    \node[draw,circle,fill=green!90,inner sep=2pt] at (2,2) {}; 

    
     \draw[line width=0.09cm,->,red] (2,2) -- (3,3); 
    \draw[line width=0.09cm,->,green] (2,1) -- (2,0); 
\end{tikzpicture}  
\end{minipage}
   \caption{Edge conflicts. Agents pointing to each other along an edge in opposite direction swap their shortest paths tables and goal agendas. }
   
    \label{fig:edge}
\end{figure}

\begin{itemize} 
\item \emph{Edge-conflicts}: Under standard MAPF rules, two agents must not traverse an edge in opposite directions in the same time step. In this case, we swap the targets \emph{and} the directions of the agents. Therefore, edge conflicts between agents can be handled immediately by swapping the target information of the agents (see Fig.~\ref{fig:edge}). 
\item For \emph{node conflicts between two approaching agents}, some conflicts with agents' shortest paths pointing to each other can be dealt with immediately (see Fig.~\ref{fig:agenda}), while other such conflicts can also incur agents' delay (see Fig.~\ref{fig:zipping} and). This delay has to be propagated recursively to
other agents along the agent-dependency graph, namely to all agents that have the location of the waiting agent as targets. For this to work, the inverted agent dependency tree has to be traversed. When marking reached nodes as visited, this backward traversal is constant-time for each node in the tree which may be traversing to neighboring cells or jumping via moving along shortest path trails through other agents located at goals. When node conflicts are being resolved and one agents is selected to wait, because it heads into the same direction on a common edge, all agents along the agent dependency tree of this agent have to wait. This dependency trees are traversed before being updated in the same iteration. This waiting will increase the time per affected agent by at most one waiting step per iteration. 
By alternating the nodes in a conflict, no agent wait forever, so that the progress on the shortest paths on the solution graph is guaranteed. By the polynomial size of the union of all shortest paths, the total number of iterations for the simulation remains polynomially bounded. 
\item
If we encounter a \emph{node conflict of several agents}, e.g. by paths intersecting, the above strategies may not suffice to resolve all the conflicts, so that, similar to the continuous case, we have to apply the assignment problem solver locally for all agents participating in the conflict to resolve it in polynomial time. We break ties by interleaving alias zipping (see Fig.~\ref{fig:zipping}), so that we warrant that some agents move in each iteration. 
\end{itemize}


Secondly, agents might have arrived at a destination and stopped. This might lead to the blocking of other agents. We resolve this problem with a concept that we call cuckoo'ing (see Fig.~\ref{fig:cuckooing}): the agents residing at the goal are pushed out of their location with the agents entering it. This is assisted by exchanging the goal and shortest-path information among them. This step also has to be executed recursively on the shortest path of an agent to move until one agent leaves the sequence of goals or all agents arrive at their desired spot. 
If there is an agent pointing to the location at the end of the cuckoo'ing process, it has to wait, and so do the agents pointing to it, so that these goal conflicts are handled prior the agent conflicts. As a positive side effect, cuckoo'ing may reduce the makespan and thus the number of iterations for the simulation. 
\hfill $\Box$ \\

\begin{figure}[t]
    \centering
\begin{small}
\begin{minipage}{3.5cm} 
\begin{tikzpicture}[scale=1]
    \draw[line width=0.04cm] (1,0) grid (3,3);
    
    \node[draw,star,fill=red!90,inner sep=3pt] at (2,1) {}; 
    \node[draw,star,fill=red!90,inner sep=3pt] at (3,2) {}; 

    \node[draw,circle,fill=yellow!90,inner sep=2pt] at (3,2) {}; 
    \node[draw,circle,fill=blue!90,inner sep=2pt] at (2,1) {}; 
    \node[draw,circle,fill=purple!90,inner sep=2pt] at (3,3) {}; 
    
    \node[above right] at (1,1) {Agent 1};
    \node[above] at (3,3) {Agent 2};
    \node[above] at (3,2) {Agent 3};
    \node[below right] at (2,1) {Goal};
    \node[below right] at (3,2) {Goal};
    
    \draw[line width=0.09cm,->,purple] (3,3) -- (3,2) -- (2,1) -- (1,0); 
\end{tikzpicture}  
\end{minipage}
\begin{minipage}{2cm} 
\begin{tikzpicture}[scale=1]
    \draw[line width=0.04cm] (1,0) grid (3,3);
    
    \node[draw,star,fill=red!90,inner sep=3pt] at (2,1) {}; 
    \node[draw,star,fill=red!90,inner sep=3pt] at (3,2) {}; 

    \node[draw,circle,fill=yellow!90,inner sep=2pt] at (2,1) {}; 
    \node[draw,circle,fill=blue!90,inner sep=2pt] at (1,0) {}; 
    \node[draw,circle,fill=purple!90,inner sep=2pt] at (3,2) {}; 
    
    \node[above right] at (0,0) {Agent 1};
    \node[above] at (3,2) {Agent 2};
    \node[above] at (2,1) {Agent 3};
    \node[below right] at (2,1) {Goal};
    \node[below right] at (3,2) {Goal};
    
\end{tikzpicture}   
\end{minipage}
\end{small}    
   \caption{Cuckoo'ing: pushing agents that have arrived at goal from their place. Agent 1 will continue with purple shortest path of Agent 2. This procedure adds shortcuts to the agent dependency graph and the interaction has to be implemented with care. }
  
    \label{fig:cuckooing}
\end{figure}
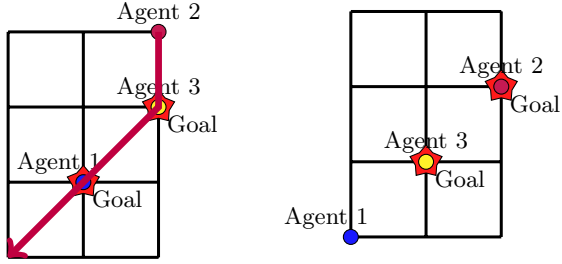

\begin{figure}[t]
    \centering
\begin{minipage}{3cm} 
\begin{tikzpicture}[scale=1]
    \draw[line width=0.04cm] (1,0) grid (3,3);

    \node[draw,circle,fill=red!90,inner sep=2pt] at (1,2) {}; 
    \node[draw,circle,fill=green!90,inner sep=2pt] at (2,3) {}; 
    \node[draw,circle,fill=yellow!90,inner sep=2pt] at (2,1) {}; 
    \node[draw,circle,fill=blue!90,inner sep=2pt] at (2,2) {}; 
    \node[draw,circle,fill=purple!90,inner sep=2pt] at (3,3) {}; 
    \node[draw,circle,fill=orange!90,inner sep=2pt] at (1,1) {}; 
    \node[draw,circle,fill=cyan!90,inner sep=2pt] at (3,2) {}; 
    
    
    \draw[line width=0.09cm,->,yellow] (2,1) -- (1,0) ; 
    \draw[line width=0.09cm,->,red] (1,2) -- (1,1); 
    \draw[line width=0.09cm,->,green] (2,3) -- (2,2); 
    \draw[line width=0.09cm,->,blue] (2,2) -- (1,1); 
    \draw[line width=0.09cm,->,orange] (1,1) -- (1,0); 
    \draw[line width=0.09cm,->,purple] (3,3) -- (3,2); 
    \draw[line width=0.09cm,->,cyan] (3,2) -- (2,1); 
\end{tikzpicture}  
\end{minipage}
\begin{minipage}{3cm}  
\begin{tikzpicture}[scale=1]
    \draw[line width=0.04cm] (1,0) grid (3,3);

    \node[draw,circle,fill=red!90,inner sep=2pt] at (1,2) {}; 
    \node[draw,circle,fill=green!90,inner sep=2pt] at (2,3) {}; 
    \node[draw,circle,fill=yellow!90,inner sep=2pt] at (1,0) {}; 
    \node[draw,circle,fill=blue!90,inner sep=2pt] at (2,2) {}; 
    \node[draw,circle,fill=purple!90,inner sep=2pt] at (3,2) {}; 
    \node[draw,circle,fill=orange!90,inner sep=2pt] at (1,1) {}; 
    \node[draw,circle,fill=cyan!90,inner sep=2pt] at (2,1) {}; 
    
    
    \draw[line width=0.09cm,->,cyan] (2,1) -- (1,0) ; 
    \draw[line width=0.09cm,->,red] (1,2) -- (1,1); 
    \draw[line width=0.09cm,->,green] (2,3) -- (2,2); 
    \draw[line width=0.09cm,->,blue] (2,2) -- (1,1); 
    \draw[line width=0.09cm,->,orange] (1,1) -- (1,0); 
    \draw[line width=0.09cm,->,purple] (3,2) -- (2,1); 
\end{tikzpicture}  
\end{minipage}
\begin{minipage}{2cm}  
\begin{tikzpicture}[scale=1]
    \draw[line width=0.04cm] (1,0) grid (3,3);

    \node[draw,circle,fill=red!90,inner sep=2pt] at (1,2) {}; 
    \node[draw,circle,fill=green!90,inner sep=2pt] at (2,2) {}; 
    \node[draw,circle,fill=blue!90,inner sep=2pt] at (1,1) {}; 
    \node[draw,circle,fill=purple!90,inner sep=2pt] at (3,2) {}; 
    \node[draw,circle,fill=orange!90,inner sep=2pt] at (1,0) {}; 
    \node[draw,circle,fill=cyan!90,inner sep=2pt] at (2,1) {}; 
    
    
    \draw[line width=0.09cm,->,cyan] (2,1) -- (1,0) ; 
    \draw[line width=0.09cm,->,red] (1,2) -- (1,1); 
    \draw[line width=0.09cm,->,green] (2,2) -- (1,1); 
    \draw[line width=0.09cm,->,blue] (1,1) -- (1,0); 
    \draw[line width=0.09cm,->,purple] (3,2) -- (2,1); 

\end{tikzpicture}  
\end{minipage} 
   \caption{Zipping: For node conflicts of agents pointing in the same direction the agents of one agent dependency subtree have to wait. The other branches are dealt with in next iteration.}
  
    \label{fig:zipping}
\end{figure}
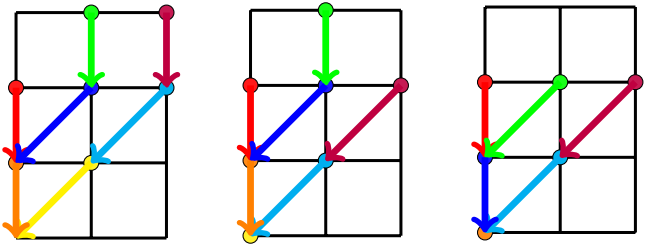

Are the solutions sum-of-cost or makespan-optimal? We first solve the assignment problem (either makespan- or sum-of-cost-optimal), and then the resulting Regular MAPF. For Regular MAPF we cannot expect optimal solutions with our polynomial-time algorithm. This can easily be validated by casting the well-known $(n^2-1)$-puzzle as a Regular MAPF, which is solvable in polynomial time, but computing an optimal solution is NP-hard~\cite{SLANEY2001119}. As we allow reassignments of goals 
during the solution process, our algorithm may not solve the Regular MAPF (with its fixed assignment), so that the sliding-tile puzzle does not serve as a counterexample. In fact, solving AMAPF optimally is polynomial~\cite{Zain}, and our algorithm finds optimal solutions in many cases (e.g., when no conflict urges agents to wait). 

\subsection{Combinatorial MAPF}

We start with single-agent CMAPF on an undirected weighted graph of $n$ nodes, $e$ edges with start node $s$ and goals $g_1,\ldots,g_m$. In the connected part of the undirected graph that contains all the goal nodes and the initial state the shortest path exists between every two 
goal nodes. Otherwise, the problem is unsolvable.

The traveling salesman problem (TSP) is closely related to CMAPF with one agent, and TSP is NP-complete even for undirected and uniformly weighted graphs \cite{tspbook}. Algorithms for solving the TSP are exponential in the number of goals (not in the graph size), as the shortest-path reduction above can be applied. For the still NP-hard Euclidean TSPs there are polytime 1.5-approximation schemes~\cite{Chr76}.
NP-hardness persists for the TSP  without edge weights, and even in general planar grid graphs~\cite{Itai}.
So why can multi-goal multi-agent path finding problems be solved in polynomial time? The key is that we allow re-targeting and do not impose exclusive visits to nodes. Instead, we solve a series of single-source shortest-path problems for a given order of visits.

\begin{theorem} [Polytime Single-Agent CMAPF]
The single-agent CMAPF problem in an undirected weighted graph with $n$ nodes, a single start node, and $m$ goal nodes can be solved in polynomial time in $m$ and $n$.    
\end{theorem}

{\bf Proof.} We assume that the subgraph containing the goals and the initial state is connected (otherwise the single-agent CMAPF problem is simply unsolvable) so that every node in this subgraph can reach each other. We now take any ordered assignment of goals to the agent, such as $(g_1,\ldots,g_m)$, and visit all goals traversing the shortest paths between every two consecutive goals $(g_i,g_{i+1})$ for $i \in \{1,\ldots,m-1\}$. \hfill $\Box$ \\


For several agents, can use any tour assignment of goals to the agents and resolve the conflicts joint single-source shortest paths traversals in succession.

\begin{definition}[Discrete Combinatorial MAPF]
Let $G=(V,E,w)$ be a weighted graph with edge cost function $w: E \rightarrow \mathbb{N}$ representing travel time.  Edge weights must be polynomially bounded and we include self-loop edges for waiting at each node with cost 1. 
We assume that the graph is undirected, so that for all $(u,v) \in E$ 
we have $(v,u) \in E$ and $w(u,v) =w(v,u).$ In the 
\emph{combinatorial MAPF problem}, for 
a set of agents $R$ with $k=|R|$ and starting locations
$S \subseteq V$ and a set of goals $W \subseteq V$ with $m =|W|$  
routes $\pi_i$ have to be found for each of the agents $i \in \{1,\ldots,k\}$, 
that in total visit all the goals and minimizes a combination of 
accumulated travel time
$\sum_{i=0}^k \sum_{(u,v) \in \pi_{i}} w(u,v)$ 
(sum-of-cost) and maximum travel time 
$\max_{i=0}^k \sum_{(u,v) \in \pi_{i}} w(u,v)$ (makespan).
\end{definition}

The algorithm we proposed extends the one that found for AMAPF. 
If the graph is disconnected, we first compute the connected components in linear time to the size of the graph and solve the problem in each connected component individually. So, w.l.o.g., we assume that the undirected graph is connected.

It is also obvious that any valid assignment of (still unvisited) goals to agents induces a partition $W_1\ldots,W_k$ of the set of remaining goals $W$, i.e., $\bigcup_{i=1}^k W_i = W$ and $W_i \cap W_j = \emptyset$ for all $1\le i \neq j \le k$. 


As a first step, once again, we compute the shortest-path reduction of the graph, starting from the goals' locations. 
With Dijkstra's algorithm, this step takes polynomial time.
%
As a second step, we again compute the matrix of pairwise distances between all destination locations in polynomial time.

Next, we use an anytime vehicle routing solver for pairwise distances to find tours for each of the agents. It is important to note that any valid (open) tour assignment of goals to the agents will do, while, only for a minimized overall travel, we need an optimized assignment. Therefore, we apply a constant number of polynomial-time rollouts in Nested Rollout with Policy Adaption (NRPA)~\cite{Rosin}, where the outcome of every rollout (see Fig.~\ref{fig:rollout}) results in a valid (open) tour assignment for all agents. When progressing to another starting position, we initiatate a new tour for the next agent, start a new timer, and skip counting the last edge. In this way, we compute valid (open)
tours in which agents are not required to return home to their starting positions.


After the agents were assigned to tours, the actual simulation starts, which resolves remaining conflicts on-the-fly and dynamically changes the goal assignments for the agents. If a goal is reached, the corresponding agent continues with the next one on its agenda, so that at each point in time, every agent has its remaining goal agenda and the next goal to visit on top, for which it can retrieve the shortest path to navigate through the graph. 
Again, we cover the subtlety of agents having already arrived at their goal location, which we resolve via cuckoo'ing. Agent-agent interactions are even rarer and are resolved with the same strategy of waiting on same-direction node conflicts 
and exchange agendas on 
opposite-direction node conflicts (Fig.~\ref{fig:agenda}), and edge conflicts (Fig.~\ref{fig:edge}). 

In between arriving at goals, therefore, we have reduced the CMAPF problem to an AMAPF problem, and mimic the simulation with
collision conflict handling from solving AMAPF, so that  
the entire simulation finishes in polynomial time. 

In summary, we obtain the following result, which adapts the goal agendas of the individual agents on-the-fly. With the NRPA algorithm, the constructed solution will have a quality-optimized solution according to the stated optimization criterion. 


\begin{theorem} [Polynomial-Time  CMAPF]
Let $m$ be the number of goals, $n$ be the number of nodes on the graph, and $k \le m$ be the number of agents. The CMAPF problem can be solved in polynomial time in $n$, $k$, and $m$. 

\end{theorem}






As solving the CMAPF problem optimally is computationally hard, for a solvable instance, we will find an approximate solution without collision. For the initial assignments of goals to the agents we use the NRPA algorithm, which is an anytime solution and can be controlled by its parameters: the number of iterations (the width of the recursion) and the level of the search (the depth of the recursion tree). As every rollout yields an assignment of goals to agents the algorithm is anytime, and warrants an goal agenda assignment after a polynomial time steps.  
The implementation of the recursive search procedure for shortest-path reduced CMAPF is provided in Figure~\ref{fig:nrpa}.
\begin{figure}[t]
    \centering  
\begin{small}
\begin{tabbing} 
\quad\=\quad\=\quad\=\quad\=\quad\=\kill
{\sc Search}(level,iteration)  \\
  \>best.score $\leftarrow \infty$; \\
  \>{\bf if} (level = 0)  \\
  \>\> best.score $\leftarrow$ {\sc Rollout}(); \\
  \>\> for (j = 0..m+k-1) best.tour[j] $\leftarrow$ tour[j];  \\
  \> {\bf else}  \\
  \>\> policy[level] $\leftarrow$ global; \\
  \>\> {\bf for} (i=0..iteration-1)  \\
  \>\>\> r $\leftarrow$ {\sc Search}(level - 1,iteration); \\
  \>\>\> {\bf if} (r.score < best.score)  \\
        \>\>\>\> best.score $\leftarrow$ r.score; \\
        \>\>\>\> for (j = 0..m+k-1) best.tour[j] $\leftarrow$ r.tour[j];  \\
        \>\>\>\> {\sc Adapt}(best.tour,level); \\
  \>\>global $\leftarrow$ policy[level];  \\
  \> {\bf return} best;  \\
\end{tabbing}
\end{small}
\caption{Main search procedure NRPA for computing initial assignments in multi-goal multi-agent search. The parameter level is counted down, while the iteration width is kept constant. The policy is kept in each level and refreshed when going down the recursion. }
 
\label{fig:nrpa}
\end{figure} 
\begin{figure}[t]
    \centering  
\begin{small}
\begin{tabbing}
\quad\=\quad\=\quad\=\quad\=\quad\=\kill
{\sc Adapt}(tour, level) \\
  \> {\bf for} (j=1..m) visits[j] $\leftarrow$ 1; \\
  \> visits[0] $\leftarrow$ k-1; succs $\leftarrow$ 0; node $\leftarrow$ 0; \\
  \> {\bf for} (j = 0..m+k-1)  \\ 
    \>\> succs $\leftarrow$ 0 \\
    \>\> {\bf for} (i = 0..m+k-1) \\
    \>\>\>  {\bf if} (visits[i] $\neq$ 0) moves[succs] $\leftarrow$ i; succs $\uparrow$ 1; \\
    \>\> $\lambda$ $\leftarrow$ (node = 0) ? 1 : 1/k \\   
    \>\> z $\leftarrow$ 0; \\
    \>\> policy[level][node][tour[j]] $\uparrow$ $\lambda$ \\
    \>\> {\bf for} (i=0..succs-1) z $\uparrow$ $e^{global[node][moves[i]]}$; \\
    \>\> {\bf for} (i=0..succs-1) \\
    \>\>\>  policy[level][node][moves[i]] $\downarrow$ $\lambda \cdot
        e^{(global[node][moves[i]])/z}$; \\
    \>\> node $\leftarrow$ tour[j]; \\
    \>\> visits[node] $\downarrow$ 1;  \\
\end{tabbing}
\end{small}
\caption{Policy adaptation in NRPA changing the policy in a given search level based on an improved tour. We use counters for the number of visits at each node, $\lambda$ is the learning parameter, $\uparrow$ is an increase, $\downarrow$ a decrease in value. 
In each iteration, the successors are generated and stored in an array, before being processed to tune the values in the policy matrix (increase for a good selection, decrease for a bad selection).
}
 
\label{fig:adapt}
\end{figure}
The randomized simulation procedure or rollout
that is executed on each leaf of the motion tree is implemented in Figure~\ref{fig:rollout}. It computes the sum-of-costs of the individual routes taken and returns it as an evaluation. The procedure that adapts the policy and steers the random simulation is shown in Figure~\ref{fig:adapt}. The core implementation trick is to generate one tour for all of the agents in common and split it at the next obtained starting locations. When generating the combined tour, the distance from the last location of the current agent to the next is reset, and the makespan, too.

\begin{figure}[t]
    \centering 
\begin{small}
\begin{tabbing}
\quad\=\quad\=\quad\=\quad\=\quad\=\quad\=\quad\=\kill 
{\sc Rollout}() \\
\>  {\bf for} (j=0..k-1)    visits[j] $\leftarrow$  1; \\
\>  visits[0] $\leftarrow$  k-1;  tour[0] $\leftarrow$  0; 
size $\leftarrow$  1;  node $\leftarrow$  0, prev $\leftarrow$  0; \\
\>  $ms \leftarrow 0$, cost $\leftarrow$ 0; 
span $\leftarrow$ 0; \\
\>  {\bf while} (size < m+k-1) \\
\>    \>succs $\leftarrow$ 0; \\
\>    \>{\bf for} (i = 0..m+k-1)  \\
\>    \>  \>{\bf if} (visits[i] $\neq$ 0) moves[succs] = i; succs $\uparrow$ 1 \\
    \>\>{\bf for} (i=0..succs-1)  value[i] $\leftarrow$ $e^{global[node][moves[i]]}$; \\ 
    \>\>i $\leftarrow$ {\sc Roulette}(value); \\  
    \>\>prev $\leftarrow$ node; node $\leftarrow$ moves[i]; \\
    \>\> tour[size] $\leftarrow$ node; size $\uparrow$ 1; visits[node] $\downarrow$ 1; \\
    \>\> cost $\uparrow$ d[prev][node]; span $\uparrow$ d[prev][node]; \\
    \>\>{\bf for} (r = 0..k-1) \\
      \>\>\>{\bf if} (start[r] = node) \\
        \>\>\>\>cost $\downarrow$ d[prev][node]; \\
        \>\>\>\>ms $\leftarrow$ max(span,ms); span $\leftarrow$ 0; \\
        \>\>\>\> {\bf break}; \\
\>  ms $\leftarrow$ max(span,ms) \\ 
\>  {\bf return} ms + cost/k; \\
\end{tabbing}
\end{small}
\caption{Polynomial-time rollout procedure in NRPA. We use counters for the number of visits at each node, 
$d$ is the precomputed shortest path distance between two nodes,
$\lambda$ is the learning parameter (usually 1), $\uparrow$ is an increase, $\downarrow$ a decrease in value. In each iteration, the successors are generated and stored in an array, before being processed to tune the values in the policy matrix. In each iteration, the successors are generated and stored in an array, before being processed. Function Roulette determines the valid successor according to his share in the fitness value. A combination between makespan and cost is returned as objective.}
 
\label{fig:rollout}
\end{figure}

\section{Experiments}

We compiled our program under Linux (Ubuntu 13.2.0-23ubuntu4) as a subsystem of a Windows 11 Pro using gcc version 13.2.0 and ran the above algorithms on one core of AMD Ryzen 7 PRO 7940U.

\begin{figure}
    \centering
    \includegraphics[width=0.6\linewidth]{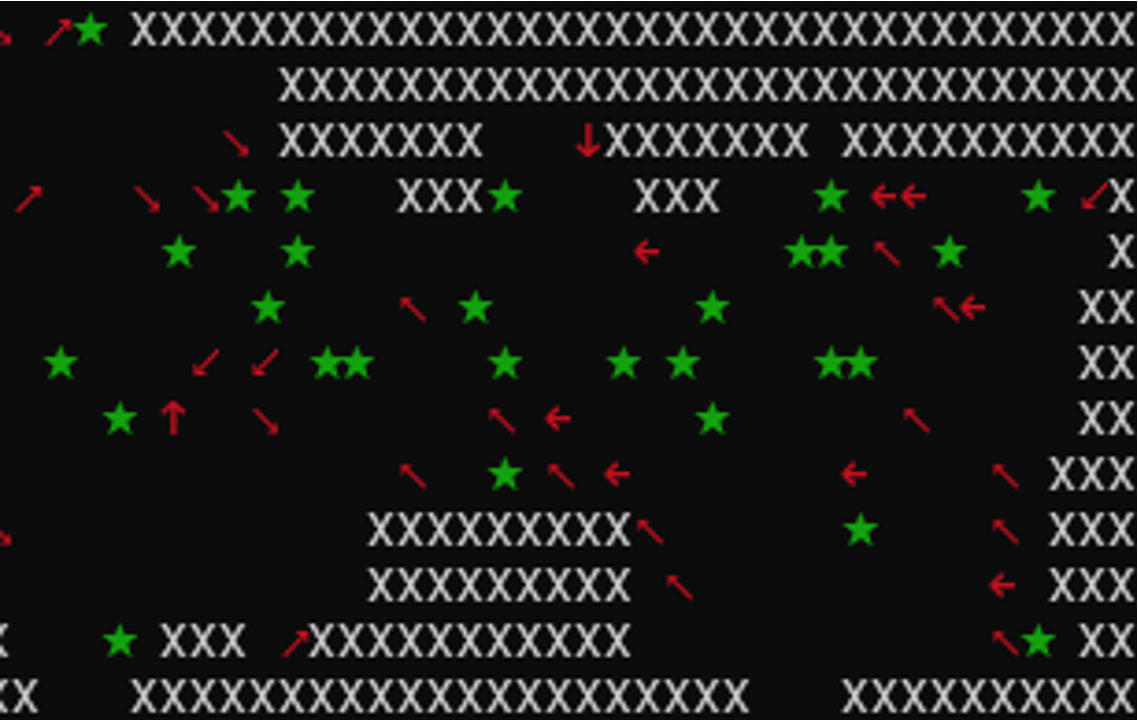}
    \caption{Visualization of multi-agent multi-goal 8-connected gridworld simulation in a terminal window with unicode characters in the cells: 
    arrows located at position of agents depict the intended direction and are the base for building the agent dependency graph; the green stars are goal cells;  obstacles/walls are marked with a white X.}
     
    \label{fig:visualiztion}
\end{figure}

\subsection{Anonymous MAPF}

For the first experiment, we selected the Boston map with 950 agents 
from the MAPF benchmark suite in Nathan Sturtevant's Moving AI Lab.  With \emph{Regular MAPF} we denote the fixed assignment given in the benchmark. With \emph{Anonymous MAPF} we denote the MAPF after reassignments of the mapping of agents to the goals. For the time being, we assume that the agents apply the ant-algorithm and exchange their goals on a potential conflict. The results are shown in Table~\ref{tab:classMAPFanonymousMAPF}.

\begin{table}[h!]
\begin{small}
\begin{tabular}{l|cc|c}
Problem & 
Sum-of-Cost &
Makspan &
Conflicts  \\ \hline
Regular MAPF &
227473 &
593  &
26420
\\
Anonymous MAPF & 
19267 &
 267&
 552 \\
 \end{tabular}
\end{small}
 \caption{Reduction of key performance indicators in a reassignment of goals by solving the assignment problem in the running example of the Boston map
 from Figure~\ref{fig:hungarian}. }
  
 \label{tab:classMAPFanonymousMAPF}
 \end{table}

We see that the makespan reduces to less than 1/2 ($45\%$), and the sum-of-cost reduces to less than 1/10 ($8.4\%$). But the most impressive impact is that the reassignment reduces the number of potential collisions in the solution simulation, where the agent exchanges their respective goals to less than 1/20 ($1.97\%$). Note that a number of conflicts that is half of the number of agents means that on average there is less than one potential collision per agent to be resolved in the simulation with cuckoo'ing and propagated delay.

\begin{figure}[htb]
    \centering
    \includegraphics[width=0.9\linewidth]{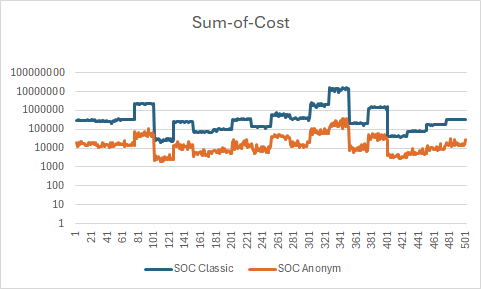}
    \caption{Comparing total travel distances of MAPF and AMAPF in the 500 benchmark instances on logarithmic scale.}
     
    \label{fig:sumofcost}
\end{figure}

\begin{figure}[htb]
    \centering
    \includegraphics[width=0.9\linewidth]{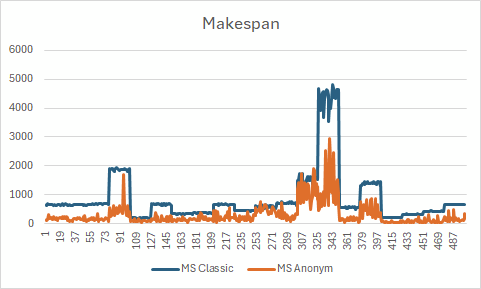}
    \caption{Comparing maximum travel distance of MAPF and AMAPF in the 500 benchmark instances.}
    
    \label{fig:makespan}
\end{figure}

\begin{figure}[t]
    \centering
    \includegraphics[width=0.9\linewidth]{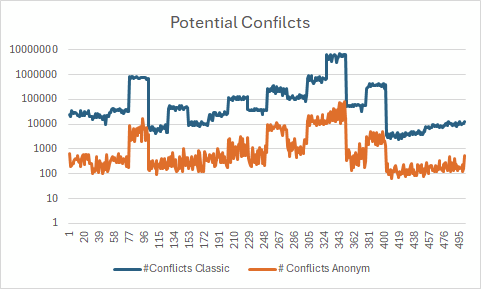}
    \caption{Comparing potential collisions of MAPF and AMAPF in the 500 benchmark instances on logarithmic scale. During simulation all conflicts are resolved on-the-fly.}
     
    \label{fig:potentialconflicts2}
\end{figure}

Across the 500 benchmark instances for the selected 20 domain in 
Figure~\ref{fig:sumofcost} shows
the reduction in sum-of-cost and Figure~\ref{fig:makespan}
the reduction in the makespan for AMAPF wrt. MAPF.  
Figure~\ref{fig:potentialconflicts2} displays the potential conflicts observed during simulation.
The gain in makespan is about $50\%$, while for the sum of cost
we estimate a reduction of one order, and for the number of potential conflicts two orders of magnitudes.

\begin{figure}[tb]
    \centering
    \includegraphics[width=0.9\linewidth]{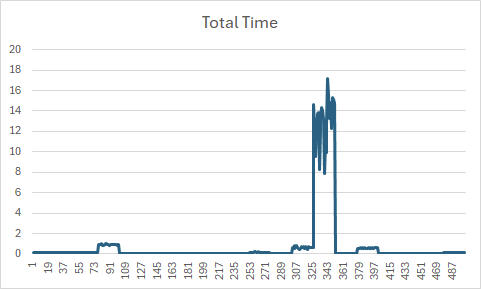}
    \caption{Total CPU Time for computing and simulating AMAP for the 500 benchmark 
    instances [in s].}
    
    \label{fig:totaltime}
\end{figure}

The running time including the simulation was mostly below 1s (see Figure~\ref{fig:totaltime}) 
and around 15 seconds for the largest maps with several thousands of agents.
If we assume that agents take more time for each
simulation step, the investment of additional efforts to calculate a better goal assignment for the agents pays off. 



\subsection{Combinatorial MAPF}

We have extended the above solution to multiple goals. The exploration of the shortest path and the subsequent distance matrix between the goals and the agents could be reused, so that individual tours for the agents were determined.

However, in this case, solutions to assignment problems are no longer sufficient, as they would lead to subtours that have to be eliminated. Instead, we are solving the vehicle routing problem, which generalizes the salesman problem. We call NRPA with a fixed number of rollouts to find an optimized solution to the tour assignment problem. In addition to the use for shortest-path navigation, the matrix for pairwise shortest paths of the goals was also used to bias the exploration of NRPA by presetting policy values.

\begin{figure}[t]
    \centering
    \includegraphics[width=0.9\linewidth]{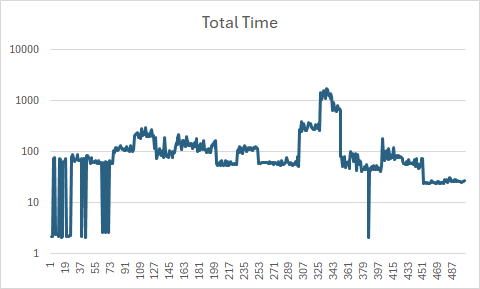}
    \caption{Total running time profile of CMAPF for the 500 benchmark instances 
    [in s] (vertical axis on logscale!). 
    Unsolvable instances are quickly found 
    by the solver and manifest as spikes in plot.}
    
    \label{fig:totaltime_CMAPF}
\end{figure}

For a number of $k$ agents we decided to let them start at the first $k$ target locations. Unfortunately, in rare cases, some goals in the domains become unreachable, so some of the problem instances were unsolvable. Fortunately, this could be detected in linear time. Using NRPA with a limited number of rollouts, we solved each of the remaining benchmark instances individually in less than half an hour. The total running times are plotted in Figure~\ref{fig:totaltime_CMAPF}.

\begin{figure}[t]
    \centering
    \includegraphics[width=0.9\linewidth]{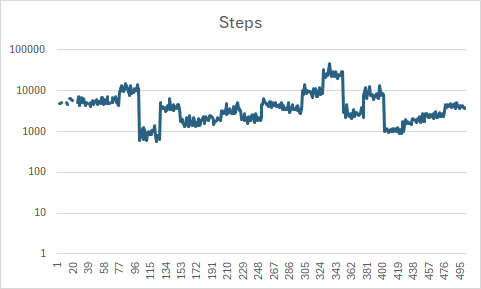}
    \caption{Number of agent steps needed to solve the problem (with unsolvable instances omitted from plot) in CMAPF for the 500 benchmark instances (vertical axis on logscale) }
    
    \label{fig:totalsteps_CMAPF}
\end{figure}

There are different trade-offs between running time and solution quality. We chose a parameterization for NRPA to conduct the entire series of experiments in one day. We validated that given more time to carry out rollouts, the solutions improve. We also checked that, as expected, the quality of the solution was generally improving for a growing number of agents.

\section{Conclusion}

Multi-agent multi-goal path finding is a fascinating problem with several applications, e.g., in indoor logistics. 
Depending on the computational resources at hand, the algorithm can be further improved by investing more time in the assignment stage.
Dynamic re-assigning of goals in the benchmark has led to a significant reduction of makespan and sum-of-cost. The number of conflicts shrinks even more drastically, 
and the remaining ones could be resolved on-the-fly.


To derive a polynomial-time algorithm, we exploited the undirectedness of the problem graph, which ensures that any greedy solution will result in valid tour assignments and goal agendas for the agents. Depending on the interpretation, multi-agent multi-goal 
path-finding leads to different complexities. Although the optimization
variant of CMAPF is NP-hard, the decision problem of finding a solution (not the one of finding a solution with cost less or equal to a given threshold) is contained in P. We describe a procedure that is theoretical as it warrants polynomial time solvability and practical as it computes 
optimized solutions to sizable benchmark problems in adequate time.
For AMAPF instances, solutions are optimal in several cases.

%
 


In the future, we will include more realistic modeling aspects. With generating solutions depth-first in rollouts, our approach extends naturally to other problem types, including pickup and delivery problem, as well as to constraints on limited resources like time windows, energy consumption, and vehicle capacity. 

\section*{Acknowledgements}
  
Thanks to Jáchym Herynek, Roman Barták, and Jiří Švancara for the discussion about this project. 

\vfill
\clearpage
\pagebreak
 
\bibliography{main}
 

\end{document}